\documentclass[12pt,a4wide]{JHEP3}
\usepackage{amsmath,amssymb}

\title{Twisted Yangian symmetry of the open Hubbard model}

\author{Alejandro De La Rosa Gomez\thanks{Email: alrg500@york.ac.uk} and Niall J. MacKay\thanks{Email: niall.mackay@york.ac.uk}\\
Department of Mathematics, University of York, York YO10 5DD, UK
}

\abstract{We show that, in the open Hubbard model with integrable boundary conditions, the bulk Yangian symmetry is broken to a twisted Yangian. We prove that the associated charges commute with the Hamiltonian and the reflection matrix, and that they form a coideal subalgebra.}

\begin{document}

\baselineskip 18pt
\parskip 10pt

\begin{section}{Introduction}

One of the classic models of condensed matter physics, the Hubbard model
\cite{Hubbard} is especially intriguing for the subtlety of the algebraic structures underpinning its integrability. A Yangian symmetry has long been known \cite{Korepin}, but this is insufficient to fix its $R$-matrix \cite{Shastry}, which does not have the typical structure associated with Yangians and quantized affine algebras.

Significant advances in understanding have come recently from the integrability of the worldsheet scattering picture of the AdS/CFT correspondence. Remarkable results \cite{Beisert,Mitev} include the identification of the Hubbard model $R$-matrix with the centrally extended $\mathfrak{su}$(2\textbar 2) (or `AdS/CFT') $S$-matrix \cite{Beisert2,Beisert3}, and the connection between the quantum deformed Hubbard chain and the $U_q(\mathfrak{su}$(2\textbar 2)) spin chain.

However, it seems that the boundary symmetry of the Hubbard model does not appear as any simple specialization of that of such models---for example, it is not simply the rational limit of a coideal subalgebra \cite{Vidas2} of the $q$-deformed Hubbard model of \cite{Beisert}.\footnote{We should like to thank Vidas Regelskis for clarification of this point.} Further, the twisted Yangian $Y(\mathfrak{su}(2),\mathfrak{u}(1))$, a modified version of which will be constructed in this paper, does not embed naturally in the twisted Yangian $Y(\mathfrak{su}(2 \vert 2),\mathfrak{su}(1 \vert 2))$ \cite{AhnNep,Vidas} of the AdS/CFT string worldsheet boundary scattering. The full relationship between the boundary Hubbard model and the models and algebras which have emerged from the AdS/CFT picture thus remain somewhat mysterious.

In this article we set aside such top-down, purely algebraic considerations and take, instead, a bottom-up approach, beginning with the physical nature of the boundary: we consider the known integrable boundary conditions of the standard, non-supersymmetric Hubbard model and attempt to discover the hidden boundary symmetry by building the generators {\em ab initio} in the spin-chain, `electronic' representation. When a Yangian $Y(\mathfrak{g})$-symmetric 1+1D model has an open boundary whose conditions preserve integrability, any remnant $\mathfrak{h}$ of the Lie symmetry typically extends to a twisted Yangian $Y(\mathfrak{g},\mathfrak{h})$ \cite{MacKay}. Such conditions for the Hubbard model appear in the form of a boundary magnetic field or chemical potential, for which the reflection (or `$K$-')matrices have been computed \cite{SW2}. These conditions break to $\mathfrak{u}(1)$ one component of the  $\mathfrak{su}(2)\times \mathfrak{su}(2)$ Lie symmetry of the bulk model. As we shall see, this extends to a twisted Yangian, modified by a boundary term which is special to the Hubbard model.

The paper is organized as follows. First, we briefly review Yangians and the conditions for boundary symmetry-breaking to preserve a twisted Yangian, following \cite{MacKay}. Secondly, we review the Hubbard model Yangian and its integrable open boundary conditions. Finally, we derive the twisted Yangian algebra of the half-infinite open Hubbard chain which is preserved under these boundary conditions, proving three crucial properties: that it commutes with the Hamiltonian, that it forms a coideal subalgebra, and that it commutes with the known $K$-matrix. The details of these calculations are presented in an appendix.

\end{section}

\vfill
\pagebreak
\begin{section}{Twisted Yangian symmetry}

Let $\mathfrak{g}$ be a compact semisimple Lie algebra generated by $\{Q_0^a\}$, $a=1,\ldots,\,$dim$(\mathfrak{g})$ with structure constants $f_{abc}$ and coproduct
\begin{eqnarray}
\Delta: \ U\mathfrak{g} \ &\rightarrow& \ U\mathfrak{g} \otimes U\mathfrak{g} \nonumber \\
Q_0^a \ &\mapsto& \ Q_0^a \otimes 1+1 \otimes Q_0^a
\end{eqnarray}
where $U\mathfrak{g}$ is the universal enveloping algebra.

Suppose that $\mathfrak{g}$ is a symmetry of a physical theory.
In a quantum integrable 1+1D model, $\mathfrak{g}$ typically extends to a hidden Yangian $Y(\mathfrak{g})$ symmetry. $Y(\mathfrak{g})$ is the enveloping algebra generated by $\{Q^a_0, Q_1^a\}$, where the $Q_1^a$ form a second set of generators of $\mathfrak{g}$ in the adjoint representation, so that
\begin{equation}\label{relation1}
[Q_0^a,Q_1^b]=f_{\;\;\;\; c}^{ab}Q_1^c
\end{equation}
with coproduct $\Delta Q_1^a = Q_1^a \otimes 1+1 \otimes Q_1^a + \frac{1}{2} f^a_{\;\;bc}Q_0^c\otimes Q_0^b$ and obeying additional so-called Drinfel'd terrific relations \cite{Drinfel}.  Finite dimensional representations of $Y(\mathfrak{g})$ are realized in one-parameter families via the automorphism
\begin{eqnarray}
\psi_{\mu}: Y(\mathfrak{g}) \ &\rightarrow& \ Y(\mathfrak{g}) \nonumber \\
Q^a_0 \ &\mapsto & \ Q^a_0 \nonumber \\
Q^a_1 \ &\mapsto & \ Q^a_1+\mu Q^a_0\,.\label{psi}
\end{eqnarray}
If the model is on the half-line, the boundary will typically break $\mathfrak{g}$ to a subalgebra $\mathfrak{h}$, say. But if the boundary condition preserves integrability---typically observed through the existence of local conserved charges or of a boundary reflection `$K$-'matrix---then one expects a further remnant of the original Yangian symmetry. In such cases $\mathfrak{h}$ is invariant under a graded involution $i$ \cite{27}. One can split $\mathfrak{g}=\mathfrak{h}\oplus\mathfrak{m}$ under $i$ such that $i(\mathfrak{h})=+1$ and $i(\mathfrak{m})=-1$. Then
\begin{equation}
[\mathfrak{h},\mathfrak{h}]\subset \mathfrak{h}, \quad [\mathfrak{h},\mathfrak{m}]\subset \mathfrak{m}, \quad [\mathfrak{m},\mathfrak{m}]\subset \mathfrak{h}\,,
\end{equation}
and $(\mathfrak{g},\mathfrak{h})$ are said to form a \emph{symmetric pair}. This property, together with  orthogonality with respect to the Killing form, $\kappa(\mathfrak{h},\mathfrak{m})=0$, guarantees the \emph{coideal property}: that the coproduct of any Yangian charge $\widetilde{Q}$ preserved at the boundary must be in the tensor product of the bulk and boundary Yangian,
\begin{equation}\label{coideal}
\Delta \widetilde{Q} \in Y(\mathfrak{g})\otimes Y(\mathfrak{g},\mathfrak{h})
\end{equation}
where $Y(\mathfrak{g},\mathfrak{h})$ is generated by $\mathfrak{h}$ and a deformation of the grade-1 $\mathfrak{m}$ generators (indexed here by $p$) \cite{DMS, MacKay2}, given by
\begin{equation}\label{tYgen}
\widetilde{Q}^p_1=Q_1^p+\frac{1}{4}[C_h,Q_0^p]
\end{equation}
where $C_h$ is the Casimir operator of $\mathfrak{g}$ restricted to $\mathfrak{h}$. These deformed generators obey commutation relations analogous to (2.2) and additional so-called Drinfel'd `horrific' relations \cite{Belliard,tYangian}.

\end{section}

\begin{section}{The Hubbard model}

The Hubbard model (the definitive work is \cite{Hubbard}) is an approximate theory used in solid state physics to describe how interactions between electrons in lattices can give rise to conducting and insulating systems. The one-dimensional model is a chain with $N$ sites, in which a kinetic `hopping' term interacts with an on-site repulsive interaction through the Hamiltonian
\begin{equation}
H=-\sum_{i=1}^N\sum_{\sigma= \uparrow,\downarrow} c_{i\sigma}^{\dagger}c_{i+1\sigma}+c_{i+1\sigma}^{\dagger}c_{i\sigma}+U\sum_{i=1}^N(n_{i\uparrow}-\frac{1}{2})(n_{i\downarrow}-\frac{1}{2})
\end{equation}
where $U$ is a coupling-constant and $c_{i\sigma}^{\dagger},\,c_{i\sigma}$ are the usual fermionic creation and annihilation operators acting on site $i$ and satisfying the only nonvanishing anticommutation relation
\begin{equation}
\{c_{i\sigma}^{\dagger}c_{j\tau}\}=\delta_{\sigma\tau}\delta_{ij},
\end{equation}
so that $n_{i\sigma}=c_{i\sigma}^{\dagger}c_{i\sigma}$ is the number density operator. The model is usually taken with periodic boundary conditions, where it may be solved using the Bethe ansatz.

Our principal interest is in the [$\mathfrak{su}(2)\times \mathfrak{su}(2)']/ \ \mathbb{Z}_2$ symmetry of the model (which, via its inclusion in the $\mathfrak{su}$(2\textbar 2)$\ltimes  \mathbb{R}^2$ symmetry of worldsheet scattering in AdS/CFT \cite{Beisert2},  is the source of the renewed recent interest from the string theory community \cite{RSS, Rej} and thereby in new generalizations such as \cite{Beisert,FFR,DFFR}).
If one defines
\begin{equation}
\mathcal{E}_i^n=c_{i\uparrow}^{\dagger}c_{i+n\downarrow} \quad \mathcal{F}_i^n=c_{i\downarrow}^{\dagger}c_{i+n\uparrow} \quad \mathcal{H}_i^n=c_{i\uparrow}^{\dagger}c_{i+n\uparrow}-c_{i\downarrow}^{\dagger}c_{i+n\downarrow}
\end{equation}
where $i$ is the spin site and $n \in \mathbb{Z}$, then one of the $\mathfrak{su}(2)$ copies is generated by $\{E_0,F_0,H_0\}$
\begin{equation}
E_0=\sum_i\mathcal{E}^0_i \quad F_0=\sum_i\mathcal{F}^0_i \quad H_0=\sum_i\mathcal{H}^0_i
\end{equation}
satisfying $[H_0,E_0]=2E_0,\ [H_0,F_0]=-2F_0$ and $[E_0,F_0]=H_0$. The summation runs over all spin sites of the chain. The other $\mathfrak{su}(2)'$, generated by $\{E_0',F_0',H_0'\}$, can be obtained through the \emph{particle-hole transformation} (PHT)
\begin{equation}
U \ \mapsto \ -U, \quad c_{i \uparrow} \ \mapsto \ (-1)^{i}c^{\dagger}_{i \uparrow}\,,
\end{equation}
hence the factor of $\mathbb{Z}_2$. From now on, to avoid repetition, we shall not include this factor when the symmetry of the model is mentioned.

The $R$-matrix of the Hubbard model is written as a tensor product of two free fermion model $R$-matrices, one for each spin layer \cite{SW}. It satisfies the Yang-Baxter equation, and (partly) underlying it is a Yangian symmetry. This Yangian was constructed \cite{Korepin} for $N \rightarrow  \infty$ and, as expected,  is composed of two copies of $Y(\mathfrak{su}_2)$ related by the PHT. One of the copies is generated by $\{E_k,F_k,H_k\}_{k=0,1}$, where the grade-1 generators are given by
\begin{eqnarray}
E_1&=&\sum_i (\mathcal{E}_i^1-\mathcal{E}_i^{-1})-\frac{U}{2}\sum_{i<j}(\mathcal{E}_i^0\mathcal{H}_j^0-\mathcal{E}_j^0\mathcal{H}_i^0) \\
F_1&=&\sum_i (\mathcal{F}_i^1-\mathcal{F}_i^{-1})+\frac{U}{2}\sum_{i<j}(\mathcal{F}_i^0\mathcal{H}_j^0-\mathcal{F}_j^0\mathcal{H}_i^0) \\
H_1&=&\sum_i (\mathcal{H}_i^1-\mathcal{H}_i^{-1})+\frac{U}{2}\sum_{i<j}(\mathcal{E}_i^0\mathcal{F}_j^0-\mathcal{E}_j^0\mathcal{F}_i^0)
\end{eqnarray}

\end{section}

 \begin{section}{Twisted Yangian symmetry of the open Hubbard chain}

If we set up the Hubbard model on a half-infinite open chain, with a single boundary, then the integrable boundary conditions consist of the presence of either a boundary magnetic field or a boundary chemical potential \cite{SW2}. These can be inserted in the hamiltonian as
\begin{equation}\label{Hopen}
H_{open}=-\sum_{i=-\infty}^{N-1}\sum_{\sigma= \uparrow,\downarrow} c_{i\sigma}^{\dagger}c_{i+1\sigma}+c_{i+1\sigma}^{\dagger}c_{i\sigma}+U\sum_{i=-\infty}^N(n_{i\uparrow}
-\frac{1}{2})(n_{i\downarrow}-\frac{1}{2})-p\phi_N
\end{equation}
where $p$ is the boundary field and $\phi_N$ is $\mathcal{H}_N^0$ in the case of a boundary magnetic field and $(\mathcal{H}_N^0)'$ in the case of a boundary chemical potential. Clearly, the symmetry of the model has now been broken to either $\mathfrak{u}(1)\times \mathfrak{su}(2)'$ or $\mathfrak{su}(2)\times \mathfrak{u}(1)'$, but integrability is unaffected \cite{SW3}. This and the fact that $(\mathfrak{su}(2),\mathfrak{u}(1))$ form a symmetric pair hint at the existence of a boundary twisted Yangian symmetry, which we shall construct shortly. We will focus on the case of a boundary magnetic field; the chemical-potential case can be obtained via PHT. Specifically, the symmetry breaking is
\begin{eqnarray}
\mathfrak{su}(2)\times \mathfrak{su}(2)' \ &\rightarrow& \ \mathfrak{u}(1) \times \mathfrak{su}(2)' \nonumber \\
\{E_0,F_0,H_0,E_0',F_0',H_0'\} \ &\rightarrow& \ \{H_0, E_0',F_0',H_0'\}
\end{eqnarray}
so we expect $Y(\mathfrak{su}(2))\times Y(\mathfrak{su}(2)')$ to break to $Y(\mathfrak{su}(2),\mathfrak{u}(1)) \times Y(\mathfrak{su}(2)')$, where the boundary symmetry is the twisted Yangian $Y(\mathfrak{su}(2),\mathfrak{u}(1))$, generated by $H_0$ and two other operators
\begin{equation}
\widetilde{E}=E_1 - \frac{U}{2} E_0H_0\,,\qquad \widetilde{F}= F_1 + \frac{U}{2} F_0H_0
\end{equation}
constructed using \ref{tYgen}. These agree with the twisted Yangian found in \cite{Belliard}, and hence satisfy the horrific relations (eqn (4.84) of \cite{Belliard}).

However, there exists a subtlety in the symmetry generators which cannot be derived by studying twisted Yangians of $\mathfrak{su}(2$\textbar$2)$. For commutation with the hamiltonian, one has to make use both of the evaluation automorphism and of the freedom to make it site-dependent, by adding a boundary term. Doing so, we find (details in Appendix \ref{A1}) that
\begin{eqnarray}
\widehat{E} & = &  \tilde{E}+\mu E_0 -p\mathcal{E}_N^0\,,\label{whE}\\
\widehat{F} & = & \tilde{F}+\nu F_0 +p\mathcal{F}^0_N\,,
\end{eqnarray}
where
\begin{equation}
\mu=-p-\frac{U}{2}-\frac{1}{p}\,,\qquad \nu = p-\frac{U}{2}+\frac{1}{p}\,,
\end{equation}
 together with $H_0$, are the charges which commute with the open Hubbard
hamiltonian and are thus conserved, so that the symmetry of the open model is not quite the expected twisted Yangian subalgebra of $Y(\mathfrak{g})$.
Although these modifications leave (\ref{relation1}) invariant, the addition of the $p$-dependent term nontrivially deforms the horrific relations for $Y(\mathfrak{su}(2),\mathfrak{u}(1))$  \cite{Belliard} by a polynomial in $p$. This is due to the $Y(\mathfrak{su}(2))$ \cite{Korepin} terrific relations' not being invariant under $Q^a_1 \mapsto Q^a_1+ \lambda Q^a_{0i}$, where $i$ denotes the site in the electronic representation.

Nevertheless, these charges satisfy the coideal property,
\begin{eqnarray}
\Delta H_0 &=& H_0 \otimes 1 + 1 \otimes H_0 \nonumber \\
\Delta \widehat{E}&=&(\widetilde{E}+ \mu E_0)\otimes 1 + 1 \otimes (\widetilde{E}+ \mu E_0-p\mathcal{E}_N^0) - UE_0\otimes H_0 \nonumber \\
\Delta \widehat{F}&=&(\widetilde{F}+\nu F_0)\otimes 1 + 1 \otimes (\widetilde{F}+\nu F_0+p\mathcal{F}^0_N)+ UF_0\otimes H_0\,,
\end{eqnarray}
because the left factors do not include the $\mathcal{E}_N^0,\mathcal{F}^0_N$ terms and so are all in $Y(\mathfrak{g})$ (details in Appendix \ref{A2}). In addition, as required for $Y(\mathfrak{su}(2),\mathfrak{u}(1))$, these twisted level-1 generators satisfy  the following commutation relations with $H_0$:
\begin{eqnarray}
[H_0, \widehat{E}] & = & [H_0,E_1]-\frac{U}{2}[H_0, E_0H_0]-p[H_0,\mathcal{E}_N^0] = 2E_1-UE_0H_0-2p\mathcal{E}_N^0 = 2\widehat{E}   \qquad \quad \nonumber \\
 \label{cop}
[H_0, \widehat{F}] & = & [H_0,F_1]+\frac{U}{2}[H_0, F_0H_0]+p[H_0,\mathcal{F}_N^0] = -2F_1-UF_0H_0-2p\mathcal{F}_N^0 = -2\widehat{F} \qquad \quad
\end{eqnarray}
and these relations are preserved by $\Delta$ (Appendix \ref{A3}). Thus the modified $Y(\mathfrak{su}(2), \mathfrak{u}(1))$ generated by $\{H_0,\widehat{E},\widehat{F}\}$ is the boundary twisted Yangian of the half-infinite Hubbard chain in the presence of a magnetic field. Further, taking the $4\times4$ $\mathfrak{su}(2)$ triple given in \cite{SW2}, the reflection matrix for a boundary magnetic field at site $i=N$ commutes with the above generators (Appendix \ref{A4}).

\end{section}

\begin{section}{Concluding Remarks}

As stressed in \cite{Hubbard} (p284), the boundary Hubbard model continues to require a deeper understanding of its algebraic structure. In principle, this should be deducible from the supersymmetric structures of AdS/CFT string worldsheet scattering, but in the absence of this we hope to have provided in this paper a useful step by constructing explicitly the twisted Yangian symmetry for the basic open Hubbard model with one integrable boundary condition in the form of a magnetic field. We found that the twisted grade-1 generators include a boundary field term not observed in constructions for other models.

From the point of view of the intrinsic study of the Hubbard model, the construction of this modified twisted Yangian lays the foundation for extended study of the boundary scattering and associated bound states. The latter have been analyzed using the Bethe ansatz (see \cite{Hubbard}, Sect.\ 8.3, and \cite{SW3,AS,BF}), but the boundary $S$-matrix has hitherto been computable only via the boundary Yang-Baxter equation \cite{SW2}, and any boundary bound state scattering would have to be computed by fusion. With the boundary's hidden charges now known, the linear conservation equations may be used instead. It would also be interesting to understand fully the role of the boundary field strength and the extent of the weak$\leftrightarrow$strong duality seen in Appendix \ref{A4}.

In the context of the richness of the Hubbard model's connections with other topics in theoretical physics, and especially in AdS/CFT, our construction provides a spur to further work in various directions. First, the connections with the twisted Yangian of the $Y=0$ maximal giant graviton \cite{Vidas} and with the deformed Hubbard model \cite{Beisert,Vidas2} should be understood. Secondly, the mathematics of the connection with the boundary analogue of the tetrahedron equation should be established \cite{IK}. Thirdly, it would be interesting to search for similar constructions in related models \cite{FFR,DFFR,Gohmann}. Finally, although the boundary conditions of \cite{SW2} remain the only ones known, it would be interesting to explore whether boundary conditions of `achiral' type \cite{Vidas3}, breaking the two $\mathfrak{su}(2)$ copies into a single diagonal $\mathfrak{su}(2)$, might be possible, or (if not) why they are ruled out.

\end{section}

\renewcommand{\thesection}{\Alph{section}}
\setcounter{section}{0}

\begin{section}{Appendix}

\begin{subsection}{Obtaining  $\mu$ and $\nu$}\label{A1}
Let $\widehat{E}=E_1+\mu E_0+\frac{\alpha}{2}E_0H_0-p\,\mathcal{E}_N^0$. We will obtain the value of $\mu$ and $\alpha$ for which $\widehat{E}$ commutes with the Hamiltonian $H_{open}\equiv-K+UV-p\mathcal{H}_N^0$. (We already know that $\alpha=-U$, since it is the constant which determines the strength of the interaction, but we will see it arise naturally from this computation). We divide the computation of the full commutator into smaller components:
\begin{flalign*}
&[-K+UV, E_1]=-2\mathcal{E}_N^0&\\
&[-K, -p \mathcal{E}_N^0]=p (\mathcal{E}_{N-1}^1-\mathcal{E}_{N}^{-1})&\\
&[-K, \mu E_0+\frac{\alpha}{2} E_0H_0]=0&\\
&[UV, \mu E_0+\frac{\alpha}{2} E_0H_0-p \mathcal{E}_N^0]=0&\\
&[-p\mathcal{H}_N^0, E_1]=-p((\mathcal{E}_{N-1}^1-\mathcal{E}_{N}^{-1})+U(\mathcal{E}_N^0+\mathcal{E}_N^0H_0))& \\
&[-p\mathcal{H}_N^0, \mu E_0+\frac{\alpha}{2} E_0H_0-p \mathcal{E}_N^0]=-2p((p+\mu)\mathcal{E}_N^0+\frac{\alpha}{2}\mathcal{E}_N^0H_0)&
\end{flalign*}
Hence, for $[H_{open},\widehat{E}]=0$,
\begin{equation}
\alpha =-U, \quad \mu=-p-\frac{U}{2}-\frac{1}{p}\,,
\end{equation}
so that $\widehat{E}=E_1+\mu E_0 -p \mathcal{E}_N^0-\frac{U}{2} E_0 H_0$ is a conserved charge. Similarly, for $\widehat{F}=F_1+\nu F_0-\frac{\alpha}{2}F_0H_0+p\mathcal{F}_N^0$ to commute with $H_{open}$,  we require the same value of $\alpha$, and $\nu = p-\frac{U}{2}+\frac{1}{p}$.
\end{subsection}

\begin{subsection}{Computation of coproducts}\label{A2}
Rewrite $E_1$ as
\begin{equation}
E_1=\sum_i (\mathcal{E}_i^1-\mathcal{E}_i^{-1})-\frac{U}{2}\sum_{i,j}A_{ij}\mathcal{E}_i^0\mathcal{H}_j^0
\end{equation}
where $A_{ij}=0$ if $i=j$, $+1$ if $i<j$ and $-1$ if $j<i$. The coproduct of the first sum is trivial but that of the second is not. However, we can use a standard trick and, for any $x \in \mathbb{Z}+{1\over 2}$, split such a sum into left ($i<x$ and $j<x$) and right ($i>x$ and $j>x$) factors, interpreting these as left and right factors of the coproduct accordingly. The boundary terms $\mathcal{E}_N^0$ and $\mathcal{F}_N^0$ therefore appear in the right factor only. For the quadratic term,
\begin{equation}
\sum_{i,j}A_{ij}\mathcal{E}_i^0\mathcal{H}_j^0 = (\sum_{\substack{ \i <x\\j<x}}+\sum_{\substack{i>x\\j>x}})A_{ij}\mathcal{E}_i^0\mathcal{H}_j^0+\sum_{\substack{i<x\\j>x}}\mathcal{E}_i^0\mathcal{H}_j^0-\sum_{\substack{i>x\\j<x}}\mathcal{E}_i^0\mathcal{H}_j^0
\end{equation}
Hence
\begin{eqnarray}
\Delta (\sum_{i,j}A_{ij}\mathcal{E}_i^0\mathcal{H}_j^0)&=&\sum_{i,j}A_{ij}\mathcal{E}_i^0\mathcal{H}_j^0 \otimes 1+1\otimes \sum_{i,j}A_{ij}\mathcal{E}_i^0\mathcal{H}_j^0 \nonumber \\ &+& E_0\otimes H_0-H_0\otimes E_0
\end{eqnarray}
and finally,
\begin{eqnarray}
\Delta \widetilde{E} &=& \Delta E_1 - \frac{U}{2}\Delta E_0 \Delta H_0 \nonumber \\
&=& E_1\otimes 1+1\otimes E_1 - \frac{U}{2}(E_0\otimes H_0-H_0\otimes E_0) \nonumber \\
&-& \frac{U}{2}(E_0H_0\otimes 1+1\otimes E_0H_0 + E_0\otimes H_0 + H_0\otimes E_0) \nonumber \\
&=& \widetilde{E}\otimes 1+1\otimes \widetilde{E}-UE_0\otimes H_0\,.
\end{eqnarray}
Note that for $\Delta \widehat{E}$ one must add $\mu E_0$ to the trivial part of the coproduct and the boundary term as specified earlier. We obtain $\Delta \widetilde{F}$ similarly, by replacing $(E,-p,-U)$ with $(F,p,U)$.
\end{subsection}

\begin{subsection}{$\Delta$ is a homomorphism of $Y(\mathfrak{su}(2),\mathfrak{u}(1))$} \label{A3}
We will check that $[\Delta H_0,\Delta \widehat{E}]=2\Delta \widehat{E}$ using \ref{cop}:
\begin{eqnarray}
[\Delta H_0, \Delta \widehat{E}] \nonumber &=& \nonumber[H_0 \otimes 1+1 \otimes H_0, (\widetilde{E}+ \mu E_0)\otimes 1 + 1 \otimes (\widetilde{E}+ \mu E_0-p\mathcal{E}_N^0) - UE_0\otimes H_0] \\ \nonumber &=& \nonumber([H_0,\widehat{E}]+\mu[H_0,E_0])\otimes 1+1\otimes([H_0,\widehat{E}]+\mu[H_0,E_0]-p[H_0,\mathcal{E}])\\ \nonumber && \nonumber - \ U[H_0,E_0]\otimes H_0 \\ \nonumber&=& \nonumber 2((\widetilde{E}+ \mu E_0)\otimes 1 + 1 \otimes (\widetilde{E}+ \mu E_0-p\mathcal{E}_N^0))-2E_0\otimes H_0\\ \nonumber &=&\nonumber 2\Delta \widehat{E}
\end{eqnarray}

\end{subsection}

\begin{subsection}{$Y(\mathfrak{su}(2), \mathfrak{u}(1))$ as a symmetry of the $K$-matrix}\label{A4}
The $K$-matrix $K^{(b)}_N(\theta,p)$ obtained in \cite{SW2} for a boundary magnetic field at site $i=N$ takes the diagonal form
\begin{equation}
\begin{pmatrix}
x_1(\theta)&0&0&0\\0&x_2(\theta)&0&0\\0&0&x_3(\theta)&0\\0&0&0&x_4(\theta)
\end{pmatrix}
\end{equation}
where
\begin{eqnarray}
x_1(\theta)&=& x_4(\theta) = (p+e^{2h}\tan\theta)(p-e^{2h}\tan\theta)\\
x_2(\theta)&=& (p+e^{2h}\tan\theta)(p+e^{-2h}\tan\theta)\\
x_3(\theta)&=& (p-e^{2h}\tan\theta)(p-e^{-2h}\tan\theta)
\end{eqnarray}
and $h(\theta)$ is defined via $\sinh 2h= \frac{U}{4}\sin2\theta$. If one takes the $4 \times 4$ representation of $H_0$ given in \cite{SW2}, then the evaluation representation $\rho$ of the $\mathfrak{su}(2)$ triple becomes
\begin{eqnarray}
\rho(H_0) &=& h \otimes 1-1 \otimes h\\
\rho(E_0) &=& e \otimes 1+1 \otimes f\\
\rho(F_0) &=& f \otimes 1+1 \otimes e
\end{eqnarray}
where $\{e,f,h\}$ is the standard $2 \times 2$ $\mathfrak{su}(2)$ triple. Set $\rho(E_1)=U(1-e^{4h})\rho(E_0)$, and note that this is invariant under  $\theta \mapsto \frac{\pi}{2}-\theta$ and is independent of the boundary field $p$. Now define
\begin{equation}
\rho_\theta(\widehat{E}) = \rho(E_1)-\frac{U}{2}\rho(E_0)\rho(H_0)+\mu \rho(E_0) +\frac{p}{\sin^2\theta}\rho(E_0)\,.
\end{equation}
 This satisfies the grade-1 generator commutation relations,  and with $\rho(\mathcal{E}_N^0)=-(\sin^2\theta)^{-1}\rho(E_0)$ yields a $4 \times 4$ representation of $\widehat{E}$ of (\ref{whE}). (One obtains the corresponding representation of $\widehat{F}$ similarly.)
This is a symmetry of the $K$-matrix,
\begin{equation}
K^{(b)}_N(\theta,p)\rho_\theta(\widehat{E})=\rho_{-\theta}(\widehat{E})K^{(b)}_N(\theta,p)\,.
\end{equation}
It is worth noting that $\mu$ plays no role in this, since $K$ is $\mathfrak{su}(2)$-invariant and automatically commutes with $\rho(E_0)$.

To interpret the $\theta=0$ pole of the $p$-dependent term, consider our original reason for adding $\mathcal{E}_N^0$ to the twisted Yangian generator: that the quadratic deformation alone cannot not fix the
commutativity of $E_1$ with the open Hubbard Hamiltonian,
\begin{equation}
[H_{open}, E_1-\frac{U}{2}E_0H_0] = p[E_1,H_{open}] \neq 0\,.
\end{equation}
But by adding $-p\mathcal{E}_N^0$ to the generator one obtains
\begin{equation}
[H_{open}, \widehat{E}] = p[E_1,H_{open}] + [\sum_{\sigma= \uparrow,\downarrow} c_{N-1\sigma}^{\dagger}c_{N\sigma}+c_{N\sigma}^{\dagger}c_{N-1\sigma},p\mathcal{E}_N^0] = 0\,.
\end{equation}
Thus the existence of the conserved twisted Yangian generator is due to the interplay of its $p$-dependent term with the hopping term in the Hamiltonian acting on the site $i=N$ and its neighbor $i=N-1$. This hopping action corresponds to the annihilation of a particle at one site and its re-creation at a neighboring site, interpreted as  motion of the particle.
If such a particle  has rapidity $\theta=0$ at $i=N$ then it is static at the boundary, and the hopping term acting on site $N$ must vanish. The twisted Yangian symmetry then disappears, and  $\rho_0(\widehat{E})$ degenerates to  $\rho(E_0)$.

To conclude, note that if the boundary is at  site $i=1$, with $K$-matrices given in \cite{SW2}, one can obtain the evaluation representation of the symmetry generators for these by making use of the relation
\begin{equation}
pK_1(p, \theta) = \frac{1}{p}K_N(\frac{1}{p}, -\theta)\,.
\end{equation}
This corresponds to a weak$\leftrightarrow$strong exchange $p \leftrightarrow \frac{1}{p}$ and a reversal of the direction of the rapidity. Hence, if $J(p,\theta)$ is a symmetry of $K_N$, then $J(\frac{1}{p}, -\theta)$ is a symmetry of $K_1$.

\end{subsection}
\end{section}

\vfill
\pagebreak


\begin{thebibliography}{100}

\bibitem{Hubbard} Essler, F.H.L., Frahm, H., G\"{o}hmann, F., Kl\"{u}mper, A. and Korepin, V.E., \emph{The One-Dimensional Hubbard Model}, Cambridge University Press (2005).

\bibitem{Korepin} Uglov, D.B., and Korepin, V.E., \emph{The Yangian Symmetry of the Hubbard Model}, Phys. Lett. \textbf{A190}:238-242 (1994), arXiv:hep-th/9310158.

\bibitem{Shastry} Shastry, B. S., \emph{Exact Integrability of the One-Dimensional Hubbard Model}, Phys. Rev. Lett. \textbf{56}:2453 (1986).

\bibitem{Beisert} Beisert, N., and Koroteev P., \emph{Quantum Deformations of the One-Dimensional Hubbard Model}, J. Phys. \textbf{A41}:255204 (2008), arXiv:0802.0777 [hep-th].

\bibitem{Mitev} Vladimir, M., Staudacher, M., and Tsuboi, Z., \emph{The Tetrahedron Zamolodchikov Algebra and the AdS$_5 \times$S$^5$ S-matrix}, arXiv:hep-th/1210.2172 (2012).

\bibitem{Beisert2} Beisert, N., \emph{The $su(2|2)$ dynamic S-matrix}, Adv. Theor. Math. Phys. \textbf{12}:945 (2008), arXiv:hep-th/0511082.

\bibitem{Beisert3} Beisert, N., \emph{The analytic Bethe ansatz for a chain with centrally extended $\mathfrak{su}(2|2)$ symmetry }, J. Stat. Mech. P01017 (2007), arXiv:nlin/0610017

\bibitem{Vidas2}
de Leeuw, M., Matsumoto, T. and Regelskis, V., \emph{Co-ideal quantum affine algebra and boundary scattering of the deformed Hubbard chain}, J. Phys. \textbf{A45}:065205 (2012), arXiv:1110.4596 [math-ph].

\bibitem{AhnNep} Ahn, C. and Nepomechie, R. I., \emph{Yangian symmetry and bound states in AdS/CFT boundary scattering}, JHEP \textbf{1005}:016 (2010), arXiv:1003:3361 [hep-th].

\bibitem{Vidas} MacKay, N. and Regelskis, V., \emph{Yangian symmetry of the $Y=0$ maximal giant graviton}, JHEP \textbf{1012}:076 (2010), arXiv:1010.3761 [hep-th].

\bibitem{MacKay} Mackay, N.J., \emph{Introduction to Yangian symmetry in integrable field theory}, Int. J. Mod. Phys. \textbf{A20}:7189-7218 (2005), arXiv:hep-th/0409183.

\bibitem{SW2} Shiroishi, M., and Wadati, M., \emph{Integrable Boundary Conditions for the One-Dimensional Hubbard Model}, J. Phys. Soc. Jpn. \textbf{66}:2288-2301 (1997), arXiv:cond-mat/9708011.


\bibitem{Drinfel} Drinfeld, V.G. \emph{Hopf algebras and the quantum Yang-Baxter equation} Sov. Math. Doklady \textbf{32} 254-258 (1985).

\bibitem{27} MacKay, N.J., and Short, B.J., \emph{Boundary scattering, symmetric spaces and the principal chiral model on the half line}, Comm. Math. Phys. \textbf{233}:313-354 (2003), arXiv:hep-th/0104212.

\bibitem{DMS} Delius, G.W., Mackay, N.J. and Short, B.J., \emph{Boundary remnant of Yangian symmetry and the structure of rational reflection matrices},
Phys. Lett. \textbf{B522}:335-344 (2001), arXiv:hep-th/0109115.

\bibitem{MacKay2} MacKay, N.J., \emph{Rational K-matrices and representations of twisted Yangians}, J. Phys. \textbf{A35}:7865-7876 (2002), arXiv:math/0205155 [math.QA].

\bibitem{Belliard} Belliard, S., and Crampe, N., \emph{Coideal algebras from twisted Manin triples},Journal of Geometry and Physics 62, pp. 2009-2023 (2012), arXiv:1202.2312v3 [math.QA].

\bibitem{tYangian} Belliard, S. and Regelskis, V., \emph{Drinfel'd basis of twisted Yangians} (2014), arXiv:1401.2143.

\bibitem{RSS}
Rej, A., Serban, D. and Staudacher, M. \emph{Planar ${\cal N}= 4$ gauge theory and the Hubbard model},
JHEP \textbf{0603}:018 (2006), arXiv:hep-th/0512077.

\bibitem{Rej}
Rej, A., \emph{Review of AdS/CFT Integrability, Chapter I.3: Long-Range Spin Chains}, Lett. Math. Phys. \textbf{99}:85-102 (2012), arXiv:1012.3985 [hep-th].

\bibitem{FFR}
Feverati, G., Frappat, L. and Ragoucy, E. \emph{Universal Hubbard models with arbitrary symmetry}, J. Stat. Mech. \textbf{0904}:P04014 (2009), arXiv:0193.0190 [hep-th].

\bibitem{DFFR}
Drummond, J.M., Feverati, G., Frappat, L. and Ragoucy, E.,  \emph{Super-Hubbard models and applications},
JHEP \textbf{0705}:008 (2007), arXiv:0712.1940 [hep-th].

\bibitem{SW} Shiroishi, M., and Wadati, M., \emph{Tetrahedral Zamolodchikov Algebra Related to the Six-Vertex Free-Fermion Model and a New Solution of the Yang-Baxter Equation}, J. Phys. Soc. Jpn \textbf{64}:12 (1995).

\bibitem{SW3} Shiroishi, M. and Wadati, M., \emph{Bethe Ansatz Equation for the Hubbard Model with Boundary Fields}, J. Phys. Soc. Jpn. \textbf{66}, 1-4 (1997).
    
\bibitem{AS} Asakawa, H. and Suzuki, M., \emph{Finite-size corrections in the XXZ model and the
Hubbard model with boundary fields}, J. Phys. \textbf{A29}:225-245 (1996).

\bibitem{BF} Bed\"urftig, G. and Frahm, H., \emph{Spectrum of boundary states in the open Hubbard chain}, J. Phys. \textbf{A30}:4139-4149 (1997), arXiv:cond-mat/9702227.

\bibitem{IK}
 Isaev, A.P. and Kulish, P.P., \emph{Tetrahedron Reflection Equations}, Mod. Phys. Lett. \textbf{A12}:427 (1997), arXiv:hep-th/9702013.

\bibitem{Gohmann}
G\"ohmann, F. and Inozemtsev, V., \emph{The Yangian symmetry of the Hubbard models with variable range hopping},
Phys. Lett. \textbf{A214}:161-166 (1996), arXiv:cond-mat/9512071.

\bibitem{Vidas3}
MacKay, N. and Regelskis, V., \emph{Achiral boundaries and the twisted Yangian of the D5-brane}, JHEP \textbf{1108}:019 (2011), arXiv:1105.4128 [hep-th].






\end{thebibliography}
\end{document}